\documentstyle{europhys}

\def\And{{\rm and\ }}

\def\stars{\bigskip\centerline{***}\medskip}

\newif\ifboo \boofalse

\def\Review#1{\boofalse{\it #1},}
\def\Name#1{{\sc #1},}
\def\Vol#1{\ifboo Vol. {\bf #1}\else{\bf #1}\fi}
\def\Year#1{\ifboo #1\else(#1)\fi}
\def\Book#1{\bootrue{\it #1},}
\def\Page#1{\ifboo {\rm p. #1}\else{\rm #1}\fi}

\begin{document}
\euro{29}{6}{1-$\infty$}{1998}
\Date{1 April 1995}
\shorttitle{A. WAGNER AN H-THEOREM FOR LATTICE BOLTZMANN}

\title{An H-Theorem for the Lattice Boltzmann Approach to Hydrodynamics}
\author{Alexander J. Wagner}
\institute{Theoretical Physics, University of Oxford, 1
Keble Rd., Oxford OX1 3BN, Great Britain}
\rec{}{}
\pacs{
\Pacs{05}{20.-y}{Statistical Mechanics}
\Pacs{05}{70.Ln}{Non-equilibrium thermodynamics, irreversible processes}
\Pacs{47}{11.+j}{Computational methods in fluid dynamics}
	}

\maketitle

\begin{abstract}
The lattice Boltzmann equation can be viewed as a discretization of
the continuous Boltzmann equation. Because of this connection it has
long been speculated that lattice Boltzmann algorithms might obey an
H-theorem.  In this letter we prove that usual nine-velocity models do
not obey an H-theorem but models that do obey an H-theorem can be
constructed. We consider the general conditions a lattice Boltzmann
scheme must satisfy in order to obey an H-theorem and show why on a
lattice, unlike the continuous case, dynamics that decrease an
H-functional do not necessarily lead to a unique ground state.
\end{abstract}

\section{Introduction}
The lattice Boltzmann approach is a method for the simulation of
hydrodynamic flow that was originally developed as a model to directly
simulate the statistical average densities of lattice gas models.
However, deriving the collision term for the lattice Boltzmann model
from a lattice gas collision term unnecessarily restricts the
Boltzmann model. Early lattice Boltzmann methods also suffered from
the exclusion principle (i.e., there can be at most one particle at a
given site), leading to an anomalous prefactor in the Navier Stokes
equation that breaks Galilean invariance \cite{frisch2}.  This
constraint was removed in the linearized lattice Boltzmann model first
introduced by Higuera and co-workers \cite{higuera}, where it was
observed that the collision operator can be linearized around a local
equilibrium and need not correspond to the detailed choice of
collision rules of the lattice gas automata, provided the operator
conserves mass and momentum.

A further simplification was introduced by Qian, d'Humi\`{e}res and
Lallemand \cite{qian}, who proposed using the Bhatnagar-Gross-Krook
(BGK) approximation \cite{bhatnagar} for the collision term in the
lattice Boltzmann method. This approximation writes the collision
operator as a function of the difference between the value of the
distribution function and the equilibrium distribution function.
For a recent review on the lattice Boltzmann method see \cite{chen}.

Another interpretation of the lattice Boltzmann approach is as a
discretized version of the continuum Boltzmann equation. 
The microscopic derivation of an H-theorem has been given by Boltzmann
for the famous Boltzmann equation (see \cite{huang}). An H-theorem
states that a functional can be defined which is a strictly decreasing
function in time. For the continuous Boltzmann equation this is the
famous H-functional
\begin{equation}
H(t) = \int d{\bf x} \int d{\bf v} f({\bf x},{\bf v},t) \ln(f({\bf
x},{\bf v},t)).
\end{equation}
Boltzmann was able to prove that for his equation
$
\frac{dH(t)}{dt}\leq 0.
$
This corresponds to the second law of thermodynamics, which states
that the entropy is a monotonically increasing function in
time. Isothermal situations are often considered for lattice Boltzmann
simulations so the energy is not conserved. In this case an
H-functional will no longer correspond to the entropy, but rather to
the free energy, which has a monotonic time behaviour in
thermodynamics.

In this letter we analyse the general conditions under which a BGK
lattice Boltzmann model can obey an H-theorem. We show that lattice
Boltzmann schemes do not automatically obey an H-theorem. It is
possible, however, to define lattice Boltzmann schemes that do obey an
H-theorem. For lattice gases H\'enon proved an H-theorem if the
collision rules obey semi-detailed balance \cite{frisch2}. This letter
for the first time describes an H-theorem for lattice Boltzmann
schemes.

In the next section we will introduce a general BGK lattice Boltzmann
scheme. We then examine which general properties we can deduce for a
lattice Boltzmann scheme that obeys an H-theorem. This will lead to a
consistency condition for the equilibrium distribution. We then show
that any lattice Boltzmann scheme with an equilibrium distribution
that obeys the consistency condition will also obey an H-theorem, and
we can construct the H-functional. Lastly we show that usual
equilibrium distributions do not obey the consistency condition and
construct one example that does obey it.

\section{The BGK lattice Boltzmann scheme}
For a single component fluid the BGK lattice Boltzmann evolution
equation is
\begin{equation} \label{RTALB}
f_i({\bf x}+{\bf v}_i \Delta t, t+ \Delta t) = f_i({\bf x},t) + 
  	\frac{\Delta t}{\tau} \left(f_i^0({\bf x},t)
	- f_i({\bf x},t)\right)
\end{equation}
where ${\bf x}$ is a discrete vector to a lattice site and the ${\bf
v}_i$'s are velocity vectors. These are chosen in a way such that
${\bf v}_i \Delta t = {\bf e}_i$ is a lattice vector of the underlying
lattice.  Formally the evolution can be decomposed into two steps: the
streaming step and the collision step.  The collision step collides the
densities according to
\begin{equation}\label{collision}
f_i^c({\bf x}, t) =f_i({\bf x},t) +
\frac{\Delta t}{\tau} \left(f_i^0({\bf x},t)- f_i({\bf x},t)\right)
\end{equation}
and the streaming step moves the density on the
lattice according to
\begin{equation}\label{streaming}
f_i({\bf x}+{\bf v}_i \Delta t, t+ \Delta t) = f_i^c({\bf x},t).
\end{equation}
Combining equations (\ref{collision}) and (\ref{streaming}) recovers
the full evolution equation (\ref{RTALB}).  Many lattice Boltzmann
applications are isothermal and do not conserve energy so the
H-functional corresponds to the free energy instead of the
entropy. The case of an isothermal model and a model that conserves
energy can thus be treated equivalently for our purpose. We will point
out the differences where they arise.

The local density $n({\bf x},t)$, the local net velocity
${\bf u}({\bf x},t)$ and, 
for a thermal model, the local kinetic energy
$\epsilon({\bf x},t)$ are given by
\begin{equation} 
n         = \sum_{i=0}^{N} f_i, \;\;\;\;
n {\bf u} = \sum_{i=0}^{N} f_i {\bf v}_i\;\;\mbox{and} \;\;
n \epsilon = \sum_{i=0}^{N} f_i {|{\bf v}_i|}^2, \label{defnue}
\end{equation}
where the sum is over the $N$ velocity vectors, ${\bf v}_i$, of the
model.  In order for this approach to simulate the continuity, the
Navier-Stokes and, for a thermal model, the heat equation, the
equilibrium distribution, $f_i^0$, has to respect the conservation of
mass, momentum and energy.  To reproduce the right form of the
transport coefficients it has to respect additional constraints,
namely higher order velocity moments of the $f_i$ have to correspond
to the equivalent moments of the continuum Boltzmann distribution
(although this is often relaxed for practical applications {\it e.g.}
by restricting the method to incompressible flow).

\section{The H-Theorem}
Let us now consider whether or not the scheme described can obey an
H-theorem. If we have a functional that always decreases in time, then
the H-functional must be locally minimal if the distribution function is
the equilibrium distribution. 

To derive the general form of this H-functional we will show that the
streaming step can not change the value of the H-functional.  Let us
consider the collisionless limit ($\tau\rightarrow\infty$) where we
have only streaming. In a periodic system it follows that the time
evolution also has to be periodic for $L!$ steps if $L$ is the number
of lattice sites. Thus, if this system obeys the H-theorem, the
streaming step cannot change the H functional.

A different way of seeing that the H-functional has to be invariant
under the streaming step is to consider a system for which the H
functional has the value $H_1$ and then to perform a 
streaming step to a new system for which the H-functional has the
value $H_2$. If the evolution obeys an H-Theorem, it follows that
$H_1\ge H_2$. We then invert all the velocities. It seems reasonable to
assume that this operation should not change the value of the
H-functional. If we now perform a streaming step on the new system, we
arrive at the original system with inverted velocities, and we can
conclude $H_2\ge H_1$ and, therefore, $H_1=H_2$.

From the invariance of the H-functional under the streaming step we
can conclude that there can be no cross-terms between the densities in
the H functional. It can therefore be written as a sum of functions of
the $f_i$ separately:
\begin{equation} \label{hfunction}
H[\{f_i\}](t)=
 \sum_{l=1}^{L} \sum_{i=0}^N h_i[f_i({\bf x}_l,t)],
\end{equation}
where $L$ is the number of lattice points and $l$ is an index that
numbers all points of the lattice.

The equilibrium distribution, $f_i^0$, is the distribution that
minimizes the H-functional under the constraint that its moments have
the same values for the conserved quantities of the distribution before
the collision. We can eliminate this constraint by introducing
Lagrange multipliers into the the H-functional: 
\begin{equation}
H[\{f_i\}] = \sum_{i=0}^{N} h_i(f_i) -
a \left(\sum_{i=0}^{N} f_i-n  \right) 
 -{\bf b} \left(\sum_{i=0}^{N} f_i {\bf v}_i - n {\bf u}
\right) 
- c \left( \sum_{i=0}^{N} f_i {|{\bf v}_i|}^2-n \epsilon \right).
\label{hl}
\end{equation}
The variation of the H-functional has to vanish for the equilibrium
distribution. We therefore obtain for the variation
\begin{equation}
\delta H[\{f_i\}]
= 
\sum_{i=0}^{N}\delta f_i \left( {h_i}'(f_i) - a 
- {\bf b} {\bf v}_i - c {\bf v}_i^2 \right)
\end{equation}
Because the $\delta f_i$ are independent, the terms for all $i$ have to
vanish independently at equilibrium. This gives us the consistency
condition for the local equilibrium distribution
\begin{equation}
f_i^0 = {h_i}'^{-1}( a + {\bf b} {\bf v}_i + c{\bf v}_i^2) \label{lfeq}.
\end{equation}
This yields a unique definition for $f^0$ if the $h_i'$ are strictly
monotonic, {\it i.e.}, if the $h_i$ are convex. 
The Lagrangian multipliers are determined by the conservation constraints
\begin{equation}
n         = \sum_{i=0}^{N} f_i^0,\;\;\;
n {\bf u} = \sum_{i=0}^{N} f_i^0 {\bf v}_i\;\;
\mbox{ and, for a thermal model,}\;\;
n \epsilon = \sum_{i=0}^{N} f_i^0 |{\bf v}_i|^2.
\end{equation}
For an isothermal model we use $c=0$ in equation (\ref{hl}).

\section{Local H-Theorem}
We can now show that for the collision step (\ref{collision}) with an
equilibrium function that obeys the consistency condition (\ref{lfeq}),
the value of the H-functional decreases. Because are only concerned with
the collision term we drop the ${\bf x}$ dependence. We
then get for the time development of the local H-functional $H^\ell$
\begin{equation}
H^\ell[\{f_i\}](t+\Delta t) - H^\ell[\{f_i\}](t) \leq 0, \hspace{0.7cm}
\forall \Delta t < \tau.
\end{equation}
The constraint $\Delta t< \tau$ excludes over-relaxation, which is
often used to simulate high Reynolds number flow because the viscosity
for lattice Boltzmann schemes has a factor $\tau-(\Delta t)/2$. The
condition $\tau=(\Delta t)/2$ corresponds to a vanishing viscosity, and
one can prove that in this limit no H-theorem can exist. The case of
$(\Delta t)/2<\tau<\Delta t$ is an interesting problem that still
warrants closer investigation.

We now provide the proof for the H-theorem. It is technically
difficult to prove it for discrete time steps. As a mathematical
simplification we introduce a continuation of the definition of the
densities for continuous time in the collision step
(\ref{collision}). The continuation is chosen so that the densities
obey the conservation constraints at all times. For these continuous
time densities we can then prove an infinitesimal H-theorem over which
we integrate to obtain the exact H-theorem for discrete time.\\ {\sc
Proof}:\\ We define for real $s\in[0,\Delta t]$
\begin{equation} \label{f1}
f_i(t+s)=f_i(t)+\frac{s}{\tau} \left(f_i^0-f_i(t)\right).
\end{equation}
Observe that $f_i^0=f_i^0(n,{\bf u},\epsilon)$ is the equilibrium
distribution for all $f_i(s)$ since the conserved quantities are the
same for all $f_i(s)$.  

Now we can prove the local H-theorem using the definition of $H$
from (\ref{hl}):
\begin{eqnarray}
&&H[\{f_i\}](t+ \Delta t) - H[\{f_i\}](t) \nonumber \\
&=& \int_0^{\Delta t} ds \sum_{i=0}^N \partial_{f_i}
( h_i(f_i(t+s)) - a f_i(t+s) 
-{\bf b} f_i(t+s) {\bf v}_i 
- c f_i(t+s) {\bf v}_i^2 \nonumber \\&&
 + \frac{1}{N} (an +{\bf b} n {\bf u}+ c n \epsilon) ) 
\partial_s f_i(t+s)\nonumber \\
&\stackrel{\mbox{(\ref{f1})}}{=}& \int_0^{\Delta t} ds \sum_{i=0}^N
\left( {h_i}'(f_i(t+s)) - a 
- {\bf b} {\bf v}_i - c {\bf v}_i^2 \right)
\times \frac{1}{\tau} \left(f_i^0-f_i(t)\right) \nonumber \\
&\stackrel{\mbox{(\ref{lfeq}
)}}{=}&\int_0^{\Delta t} ds \sum_{i=0}^N
 \left({h_i}'(f_i(t+s))-{h_i}'(f_i^0)\right)
\times\frac{\alpha(s)}{\tau}(f_i^0-f_i(t+s))  \nonumber\\
&\leq& 0 \label{HProof}
\end{eqnarray}
if $h'$ is non-decreasing or, equivalently, if $h$ is convex. We have
$\alpha(s)=(1-s/\tau)^{-1}$, which is always positive because $s<\Delta t<\tau$.

\section{Global H-theorem}
The total entropy ${H}[\{f_i\}](t)$ defined in (\ref{hfunction}) is
non-increasing  at every lattice site in the collision step and is
unchanged in the streaming step. We therefore have the global H-theorem
\begin{equation}
H[\{f_i({\bf x})\}](t+\Delta t) 
	- H[\{f_i({\bf x})\}(t)], \leq 0 
\hspace{0.7cm}\forall \tau >\Delta t.
\end{equation}

\section{The Global Equilibrium Distribution}
In statistical mechanics the H-theorem is used to prove the existence
of a unique equilibrium state of the system. It will turn out that for
lattice systems this is not necessarily the case. Demanding the
existence of a well-defined ground state gives us a constraint
for the structure of the lattice. This is equivalent to the
constraint that there are no spuriously conserved quantities.

Since we have a global H-theorem and the H-functional is bounded, we
know that the scheme has to converge to some minimal value of the
H-functional. We will now examine what information about the final
state we can extract from our H-Theorem.

Since the H-Functional does not change in the final state we know
that it also cannot change locally in the collisions. Therefore the
local distributions must be equilibrium distributions 
$f_i(t) = f_i^0$.
We can conclude that for
large $t$ the system converges to a state that has local equilibrium
distributions everywhere. Furthermore, the streaming step has to
transform one state of local equilibrium into another (or the same)
state of local equilibrium.

Whether these conditions force the global equilibrium to be
homogeneous depends on the lattice and the set of velocity vectors
$\{{\bf v}_i\}$. This question is related to the problem of spurious
invariants (see, for instance, \cite{benzi} and references therein).
Spurious invariants are conserved quantities that do not correspond to
any physical quantities. In a four-velocity model on a square lattice,
for instance, the total momentum of all even and odd lattice sites is
separately conserved.  If the global equilibrium is constrained to be
homogeneous, then there cannot be any spurious invariant. If, however,
inhomogeneous final states are possible, then these state can be
characterized by at least one spuriously conserved quantity.

\section{Why conventional lattice Boltzmann schemes cannot have an
H-theorem} 
We show that assuming an H-theorem leads to a structure of the
equilibrium distribution that is different from the
usual structure of the equilibrium distribution.  The usual BGK
lattice Boltzmann schemes have a polynomial equilibrium
distribution. For an isothermal model it takes the form
\cite{qian,flekkoy,swift,enzo}
\begin{equation}\label{tradfeq}
f_i^0 = A_\sigma n + B_\sigma n u_\alpha v_{i\alpha} 
+ C_\sigma n u_\alpha u_\alpha
+ D_\sigma n u_\alpha u_\beta v_{i\alpha} v_{i\beta},
\end{equation}
where $A_\sigma, B_\sigma, C_\sigma,$ and $D_\sigma$ are constants and
$\sigma$ is an index distinguishing velocities with different
magnitudes. If this equilibrium distribution is to be derived from an
H-functional of the form given in equation (\ref{hfunction}) then the
$h'^{-1}(x)$ have to be quadratic polynomials. We obtain
\begin{eqnarray}
f_i^0 &=& {h_i}'^{-1}(a + b_\alpha v_{i\alpha})\nonumber\\
&=& \alpha_\sigma + \beta_\sigma (a+b_\alpha v_{i\alpha}) +
\gamma_\sigma (a+ b_\alpha v_{i\alpha})^2\nonumber\\
&=& (\alpha_\sigma+\beta_\sigma a+ \gamma_\sigma a^2) + (\beta_\sigma + 2
\gamma_\sigma a) b_\alpha v_{i\alpha} + \gamma_\sigma b_\alpha b_\beta
v_{i\alpha} v_{i\beta} \label{feq}
\end{eqnarray}
where the coefficients $\alpha_\sigma$, $\beta_\sigma$ and
$\gamma_\sigma$ are constants that cannot depend on $n$ or $\bf u$.
The Lagrange multipliers $a$ and $b_\alpha$ are determined by the
conservation laws. In order for the coefficients in (\ref{feq}) to be
linear in $n$, we require $\alpha_\sigma=\beta_\sigma=0$. Evaluation shows
that the resulting
coefficients are not quadratic in $u_\alpha$ for any choice of
$\gamma_\sigma$. In particular, we get
\begin{equation}
a^2=n \frac{\gamma_1+2\gamma_2+\sqrt{(\gamma_1+2\gamma_2)(-2u^2
\gamma_0+(1-8u^2)\gamma_1 +2(1-4u^2)\gamma_2)}}{2(\gamma_1+ 2\gamma_2)
(\gamma_0+ 4 (\gamma_1+ \gamma_2))}
\end{equation}
which has a more complicated $u$ dependence for all $\gamma_\sigma$
than equation (\ref{tradfeq}). All lattice Boltzmann schemes of which
we are we are aware have a polynomial $u$ dependence and therefore
cannot obey an H-theorem.

\section{A lattice Boltzmann scheme with an H-theorem}
If we use the classical choice for the H-functional
\begin{equation}
H(\{f_i\}) = \sum_l \sum_i f_i \ln(f_i)
\end{equation}
for a thermal model, i.e., a model with mass, momentum and energy
conservation, 
we get for the equilibrium equation a Maxwell-Boltzmann distribution
\begin{equation}
f_i^0 = N \exp\left( ({\bf v}_i -{\bf U})^2/T \right)
\end{equation}
where $N$, ${\bf U}$ and $T$ are the Lagrange multipliers. This scheme
simulates the continuity, Navier-Stokes and heat equations to an
approximation that depends on the choice of lattice (because the
higher order moments needed for the Chapman Enskog expansion do not
necessarily coincide for the discrete and continuum
case). Unfortunately, the Lagrange multipliers cannot be expressed
analytically in terms of the conserved quantities, but have to be found
by numerically solving the non-linear equation. For a regime where the
Navier-Stokes and heat equations are recovered, the Lagrange multipliers
are well approximated by ${\bf U} \sim {\bf u}$ and $T\sim \theta= 1/d
(\epsilon - n {\bf u}^2)$, where d is the number of spatial dimensions.

The advantage of this scheme is that it is numerically
stable for $\tau>\Delta t$. This is ensured by the H-theorem because
numerical instabilities lead to inhomogeneities that would increase
the H-functional.

\section{Conclusions}
We have shown how lattice Boltzmann models can
be constructed to obey an H-theorem and that the usual
choice of the equilibrium distribution is incompatible with an
H-theorem. For traditional schemes no H-functional can exist.

It will be interesting to investigate further examples and applications
of lattice Boltzmann methods which obey an H-theorem. One apparent
advantage of these schemes is the numerical stability that results
from the constraint which minimizes the H-functional. We believe that
constructing lattice Boltzmann schemes with H-functionals will help to
improve the stability lattice Boltzmann schemes.

\stars It is a pleasure to thank Dominique d'Humi{\`e}res and Colin
Marsh for several helpful discussions.



\end{document}